\documentclass[aps,groupedaddress,showpacs]{revtex4}
\usepackage{graphicx}
\begin{document}
\title{Dynamical properties of S=1 bond-alternating Heisenberg chains at finite temperatures}
\author{Takahumi Suzuki and Sei-ichiro Suga}
\affiliation{Department of Applied Physics, Osaka University, Suita, Osaka 565-0871, Japan}
\date{\today}
\begin{abstract}
Dynamical structure factors of the $S=1$ bond-alternating spin chains in the dimer phase are calculated at finite temperature, using the pair dynamical correlated-effective-field approximation. At $T=0$, the delta-function-type peak of the one-magnon mode appears. When temperature is increased, such a sharp peak is broadened and the additional weak peak caused by the excitation from the triplet state to the quintet state emerges in the higher energy region. The results are discussed in comparison with those obtained by the exact diagonalization method.
\end{abstract}
\maketitle
\section{Introduction}
There has been a continued interest in $S=1$ bond-alternating Heisenberg chains both theoretically and experimentally. It was shown that a quantum phase transition between the Haldane phase and the dimer phase takes place depending on the bond-alternating ratio \cite{aff,ah}. The quantum phase transition of the system was studied by including the effects of a single-ion anisotropy. The phase diagram for the parameters of the bond-alternating ratio $(\alpha)$ and the single-ion anisotropy $(D)$ was obtained \cite{tone,chen,koga}, where the bond-alternating ratio is $1:\alpha$. Along the phase boundary of the Haldane-gap and dimer phases, the system becomes gapless.

Recent progress on material synthesis has made it possible to study the elementary excitation as well as thermodynamic properties of $S=1$ bond-alternating Heisenberg chains experimentally. 
Furthermore, dynamical properties have been investigated by inelastic neutron-scattering experiments for a typical anisotropic dimer-phase compound $\rm{Ni(C_{9}D_{24}N_{4})(NO_{2})ClO_{4}}$ (abbreviated to NTENP) \cite{hagiwara}. 
It was shown that spin dynamics of NTENP in transverse magnetic fields show noticeable different features as compared with those of an anisotropic Haldane-gap compound $\rm{Ni(C_{5}D_{14}N_{2})_{2}N_{3}(PF_{6})}$ \cite{Zhe1}. 
The experimental findings have been analyzed on the basis of the numerical diagonalization calculation and explained in viewpoint of the field dependence of the excitation continuum \cite{me2}.

Synthesized compounds for $S=1$ bond-alternating Heisenberg chains so far are in the dimer phase. NTENP is located relatively close to the gapless line. Some compounds are located far away from the gapless line. 
For example, the parameters of $\rm{[Ni_2(dpt)_2(\mu-ox)(\mu-N_3)](PF_6)[dpt=bis(3-aminopropyl)amine, ox=C_2O_2]}$ (abbreviated to NDOAP) are evaluated as $\alpha=0.1$ and $D=0$ and the system is situated close to the isolated dimer system \cite{naru4}. 
Dynamical structure factors (DSF) of $S=1$ bond-alternating Heisenberg chains at zero temperature were calculated, using the continued fraction method based on the Lanczos algorithm \cite{me1,me2}. Field dependence of spectral intensities of the magnon isolated mode and the excitation continuum was investigated in the Haldane-gap and the dimer phases. However, temperature dependence of the dynamical structure factor has not yet been fully investigated. 
It may be desirable to investigate dynamical properties of $S=1$ bond-alternating Heisenberg chains at finite temperatures.

In this Letter, we calculate dynamical structure factors at finite temperatures in the dimer phase, using a pair dynamical correlated-effective-field approximation (Pair-DCEFA) \cite{kokado}. Pair-DCEFA was successfully applied to calculate the dynamical structure factor as well as the static susceptibility of the $S=1/2$ bond-alternating Heisenberg chain at finite temperatures \cite{kokado}. 
In Section 2, we derive the effective Hamiltonian and formulate the expressions for the static susceptibility and the DSF, by using Pair-DCEFA. 
In Section 3, we first show the numerical results for the DSF at zero temperature and the temperature dependence of the susceptibility. The results are compared with those obtained by numerical diagonalization method. We then show the results for the DSF at finite temperatures. Characteristics of the isolated modes are discussed. Brief summary is given in the last section.

\section{Model and method}
Let us consider the $S=1$ bond-alternating Heisenberg chain described by the following Hamiltonian, 
\begin{eqnarray}
\mathcal{H}= J \sum_{i} \Bigl( \mathbf{S}_{i,1} \cdot \mathbf{S}_{i,2} +
  \alpha \mathbf{S}_{i-1,2}\cdot\mathbf{S}_{i,1} \Bigl) \hspace{2mm} (J>0),
\label{eqn:ham1}
\end{eqnarray}
where $\mathbf{S}_{i,1}$($\mathbf{S}_{i,2}$) denotes the operator of the spin on the left side (right side) in the unit cell $i$.

In Pair-DCEFA, we first consider the singlet, triplet, and quintet states of an isolated spin pair on the strong bond. 
To take into account the effects of the interaction between spin pairs, we decouple the second term of the Hamiltonian (\ref{eqn:ham1}) and then obtain the effective Hamiltonian in Pair-DCEFA as 
\begin{eqnarray}
\mathcal{H}_{\rm eff} = J \sum_{i} \mathbf{S}_{i,1} \cdot \mathbf{S}_{i,2} 
  + \alpha J \sum_{i} 
       \Big[\mathbf{S}_{i,1} \cdot (\langle \mathbf{S}_{i-1,2} \rangle
         - \lambda \langle \mathbf {S}_{i,1}\rangle) 
         + \mathbf{S}_{i,2} \cdot (\langle \mathbf{S}_{i+1,1} \rangle - 
           \lambda \langle \mathbf {S}_{i,2}\rangle) \Big], 
\label{eff1}
\end{eqnarray}
where $\langle \mathbf{S}_{j\nu} \rangle$ $\; (\nu=1,2)$ denotes the spontaneous moment and $\lambda$ is a correlation parameter determined by the self-consistency equation derived from the fluctuation-dissipation theorem. We confine ourselves to consider the paramagnetic state. 
On the basis of the effective Hamiltonian, the self-consistency equation for $\lambda$ is obtained as 
\begin{eqnarray}
\frac{8}{3} &=& \frac{1}{N} \sum_{q} 
                \frac{J}{[\hbar \omega_+(q)]^2 - [\hbar \omega_-(q)]^2} 
     \Big\{
     \coth \left[ \frac{\beta\hbar\omega_+(q)}{2} \right]
     \left[ d_1 \hbar\omega_+(q) - \frac{d_2J^2}{\hbar\omega_+(q)} \right] 
\nonumber \\
    &-& \coth \left[ \frac{\beta\hbar\omega_-(q)}{2} \right]
    \left[ d_1 \hbar\omega_-(q) - \frac{d_2J^2}{\hbar\omega_-(q)} \right]
    \Big\}, 
\label{self}
\end{eqnarray}
where $N$ is the total number of unit cells, $\beta=1/k_{\rm B} T$, $d_1= (4/3)(2\Delta \rho_0+5\Delta \rho_1)$, and $d_2=(4/3)(8\Delta \rho_0+5\Delta \rho_1)$ with $\Delta \rho_0=\rho_{\rm s}-\rho_{\rm t}$ and $\Delta \rho_1=\rho_{\rm t}-\rho_{\rm q}$.   
Here $\rho_{\rm s}(=\exp(2\beta J)/Z)$, $\rho_{\rm t}(=\exp(\beta J)/Z)$, and $\rho_{\rm q}(=\exp(-\beta J)/Z)$ are the contributions from the singlet, triplet, and quintet states with $Z=\exp(2\beta J) + 3\exp(\beta J) + 5\exp(-\beta J)$. In thermal equilibrium, the excitation from the singlet state to the triplet state and the excitation from the triplet state to the quintet state take place. 
When $T \rightarrow 0$, we find that $\rho_{\rm s} \rightarrow 1$, $\rho_{\rm t} \rightarrow 0$, and $\rho_{\rm q} \rightarrow 0$.

The dispersion relation of the one-magnon mode is given by  
\begin{eqnarray}
\omega_{\pm}(q) &=& \frac{J}{\sqrt{2}} 
        \Bigl[\bigl( 5-\alpha d_1 
              \big\{ \lambda + \cos[q(c+c^{\prime})] \big\} \bigl) \nonumber \\    &\pm& \sqrt{
              \bigl( 5-\alpha d_1 
              \big\{ \lambda + \cos[q(c+c^{\prime})] \big\} \bigl)^2 
            - 4\bigl( 4-\alpha d_2 
              \big\{ \lambda + \cos[q(c+c^{\prime})] \big\} \bigl) 
        }
        \Bigl]^{\frac{1}{2}},
\label{pair2}
\end{eqnarray}
where $\omega_{+}(q)$ and $\omega_{-}(q)$ denote the dispersion relations of the excitations from the singlet state to the triplet state and from the triplet state to the quintet state, and $c$ and $c^{\prime}$ are the lengths between the neighboring spins along the strong bond and the weak bond, respectively.

By using $\lambda$ and $\omega_{\pm}(q)$ thus obtained, the static susceptibility $\chi(T)$ and the dynamical structure factor $S(q,\omega)$ in Pair-DCEFA are expressed as
\begin{eqnarray}
\chi(T) = \frac{2(g\mu_{\rm B})^2 (\rho_{\rm t}+5\rho_{\rm q})}
          {1+(\lambda-1)\alpha J (\rho_{\rm t}+5\rho_{\rm q})}, 
\label{sus}
\end{eqnarray}
\begin{eqnarray}
S(q,\omega) &=& \frac{2J}{1-e^{-\beta \hbar \omega}} \cdot 
     \frac{2[1-\cos(qc)]}{[\hbar \omega_+(q)]^2 - [\hbar \omega_-(q)]^2} 
     \Big\{ \frac{d_1 [\hbar \omega_+(q)]^2 - d_2 J^2} 
        {\hbar \omega_{+}(q)}{\frac{\delta}{[\omega-\omega_+(q)]^2+\delta^2}} 
        \nonumber\\
&-& \frac{d_1 [\hbar \omega_-(q)]^2 - d_2 J^2} 
{\hbar \omega_{-}(q)}{\frac{\delta}{[\omega-\omega_-(q)]^2+\delta^2}} \Big\}. 
\label{eqn:pcdfa}
\end{eqnarray}
In the following, we set $c=c^{\prime}=1$, $k_{\rm B}=\hbar=1$, and $\delta=5.0 \times 10^{-2}$. The energy is measured in units of $J$.

\section{RESULTS}
We solve Eqs. (\ref{self}) and (\ref{pair2}) numerically to obtain $\lambda$ and $\omega_{\pm}(q)$ in given $T$ for fixed $\alpha$. 
\begin{figure}[thb]
\begin{center}
\includegraphics[trim=0mm 0mm 0mm 0mm,scale=0.40,clip]{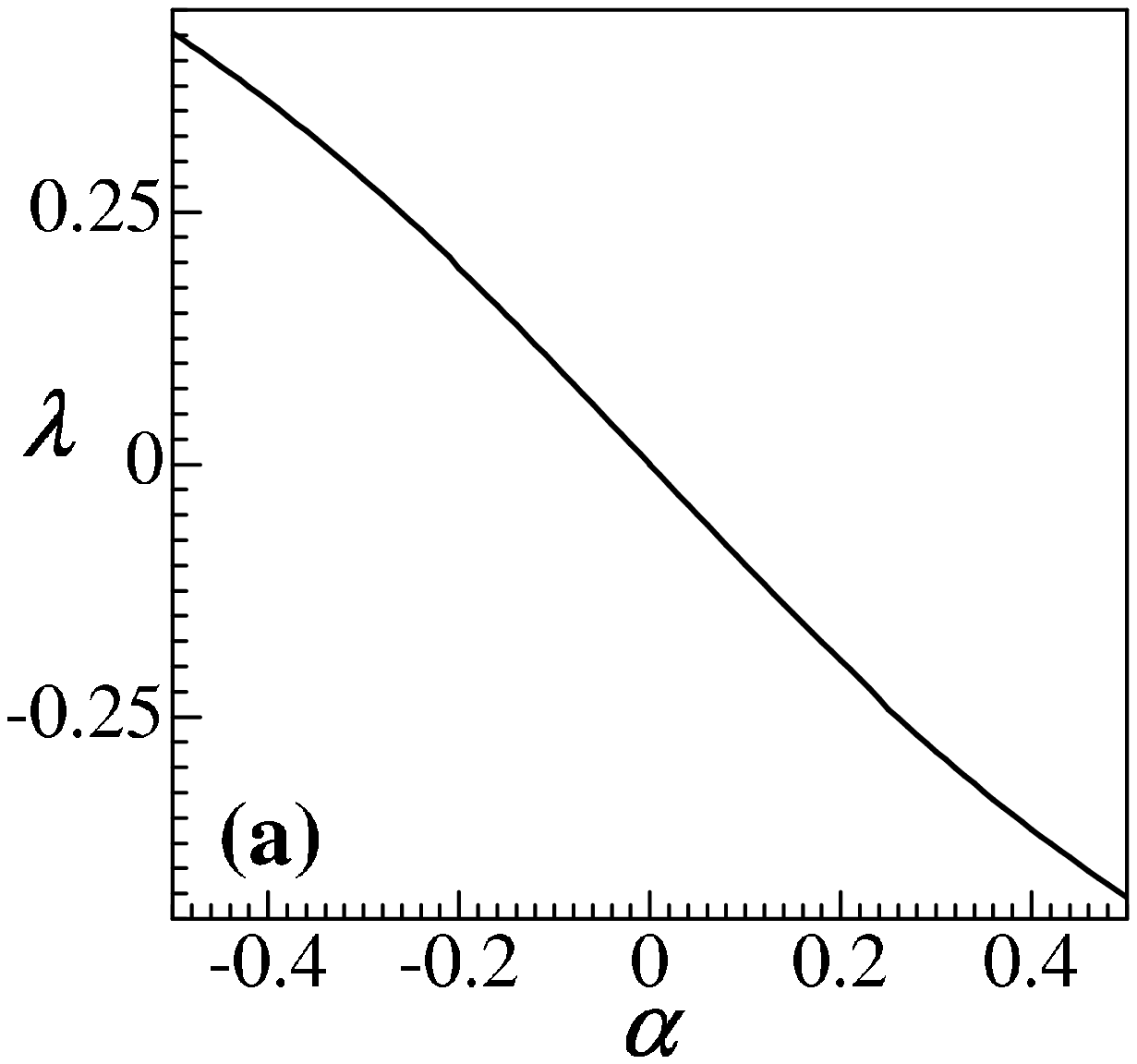}
\includegraphics[trim=0mm 0mm 0mm 0mm,scale=0.40,clip]{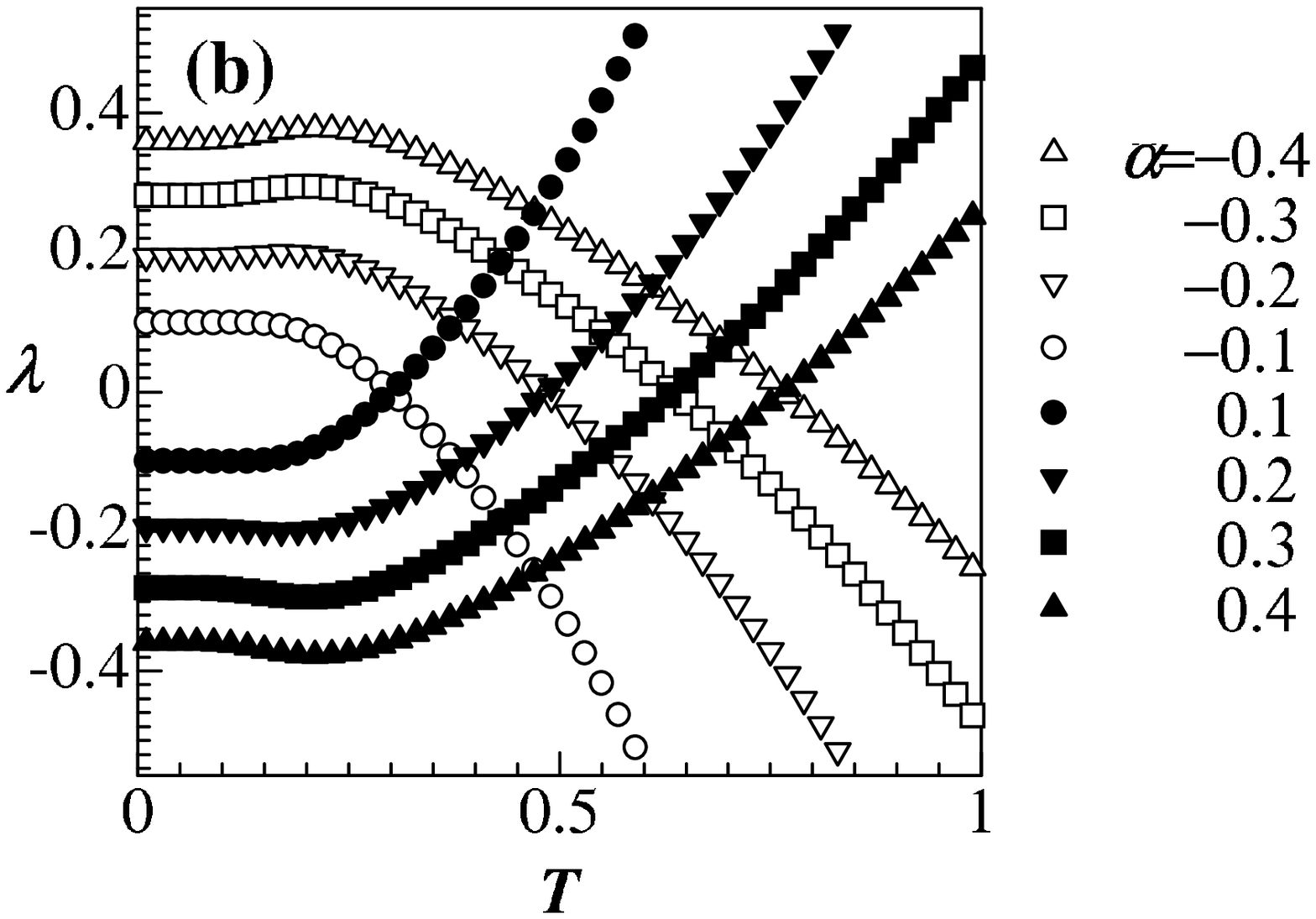}
\vspace{1mm}
\caption{(a) The correlation parameter as function of $\alpha$ at $T=0$. 
(b) The temperature dependence of the correlation parameter for several values of $\alpha$. }
\label{fig:0}
\end{center}
\end{figure}
The results for $\lambda$ are shown in Fig. \ref{fig:0}. 
In $|\alpha| > 0.6$, the convergence of the self-consistent procedure becomes worse at finite temperatures. Such ineffectiveness of the self-consistent treatment may relate to the quantum phase transition around $\alpha \sim 0.6$ \cite{singh,kato,yamamoto,totsuka,kitazawa,kohno}. 
The behavior of $\lambda$ in this system is in contrast with that in the $S=1/2$ bond-alternating spin chain, where temperature dependence of $\lambda$ is obtained even at $|\lambda|=1$ \cite{kokado}. 
The different feature of $\lambda$ may represent characteristics of spin correlations in both systems.

\begin{figure}[thb]
\begin{center}
\includegraphics[scale=0.4,clip]{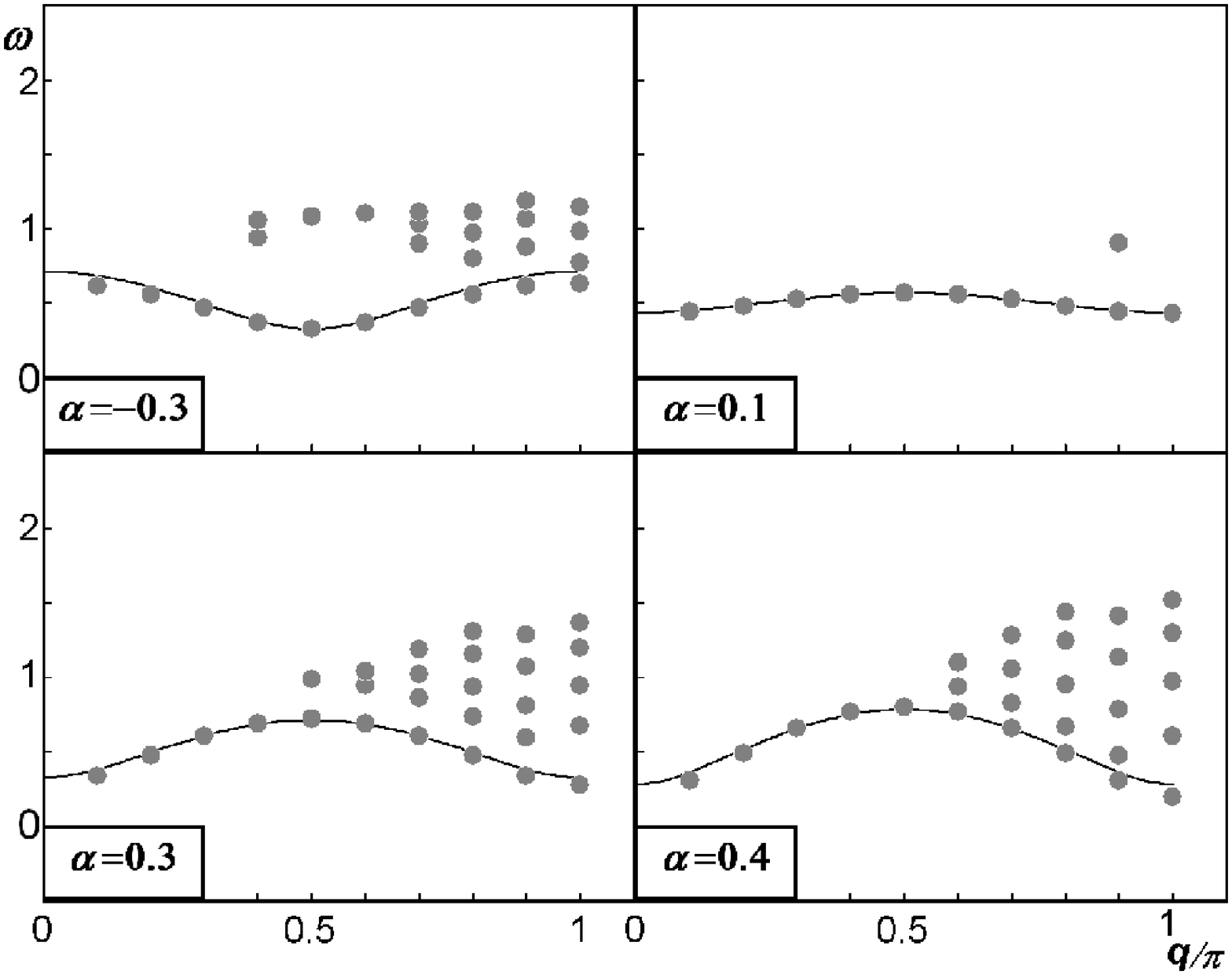}
\vspace{1mm}
\caption{The dispersion curves $\omega_-(q)$ at $T=0$ for $\alpha=-0.3, 0.1, 0.3$, and $0.4$. The results obtained by Pair-DCEFA are expressed by the solid lines and those obtained by the exact diagonalization method for 20 spin systems are plotted by dots.}
\label{fig:1-disp}
\end{center}
\end{figure}
We next show the dispersion curve $\omega_-(q)$ at $T=0$. In Fig. \ref{fig:1-disp}, the results are expressed in the extended zone scheme. 
The dispersion curves obtained by Pair-DCEFA are expressed by the solid lines for $\alpha=-0.3, 0.1, 0.3$, and $0.4$. As a reference, the excitation energies obtained by the exact diagonalization method for 20 spin systems \cite{me1} are plotted by dots.   
For $\alpha=0.1$ and $0.3$ the agreements between both results are excellent, while for $\alpha=-0.3$ and $0.4$ the deviations become noticeable in $q \sim 0$ and $\pi$. 
Thus, Pair-DCEFA may be effective in $-0.3 < \alpha \leq 0.3$ for the $S=1$ bond-alternating Heisenberg chain.

\begin{figure}[thb]
\begin{center}
\includegraphics[trim=0mm 45mm 0mm 45mm,scale=0.45,clip]{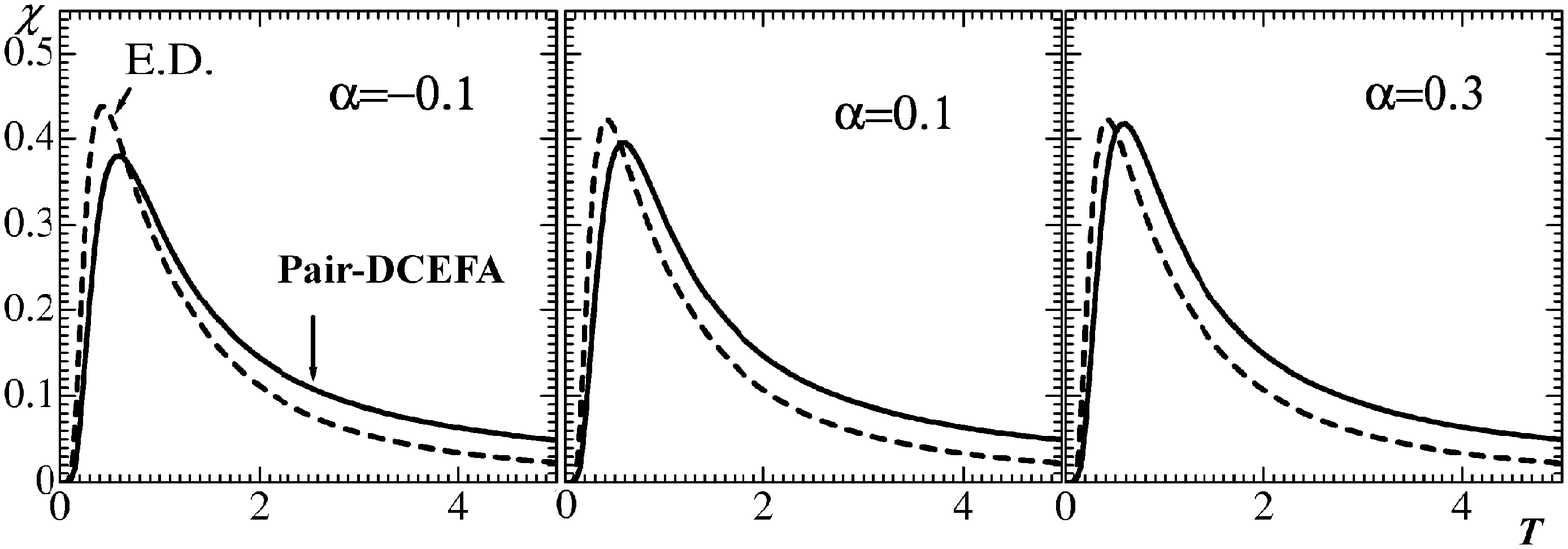}
\vspace{1mm}
\caption{The static susceptibility $\chi(T)$ for $\alpha=-0.1, 0.1$, and $0.3$. The results obtained by Pair-DCEFA are expressed by the solid lines and those obtained by the exact diagonalization (ED) method are expressed by the broken lines.}
\label{fig:2-susc}
\end{center}
\end{figure}
Temperature dependence of the static susceptibility $\chi(T)$ is shown in Fig. \ref{fig:2-susc}. 
The results for $\alpha=-0.1, 0.1$, and $0.3$ are compared with those obtained by the exact diagonalization method for 8 spin systems. 
Note that $\chi(T)$ by the exact diagonalization method exhibits little size dependence around 8 spin systems.  
The agreement between both results becomes better in low temperatures.

\begin{figure}[thb]
\begin{center}
\includegraphics[trim=0mm 0mm 0mm 0mm,scale=0.5,clip]{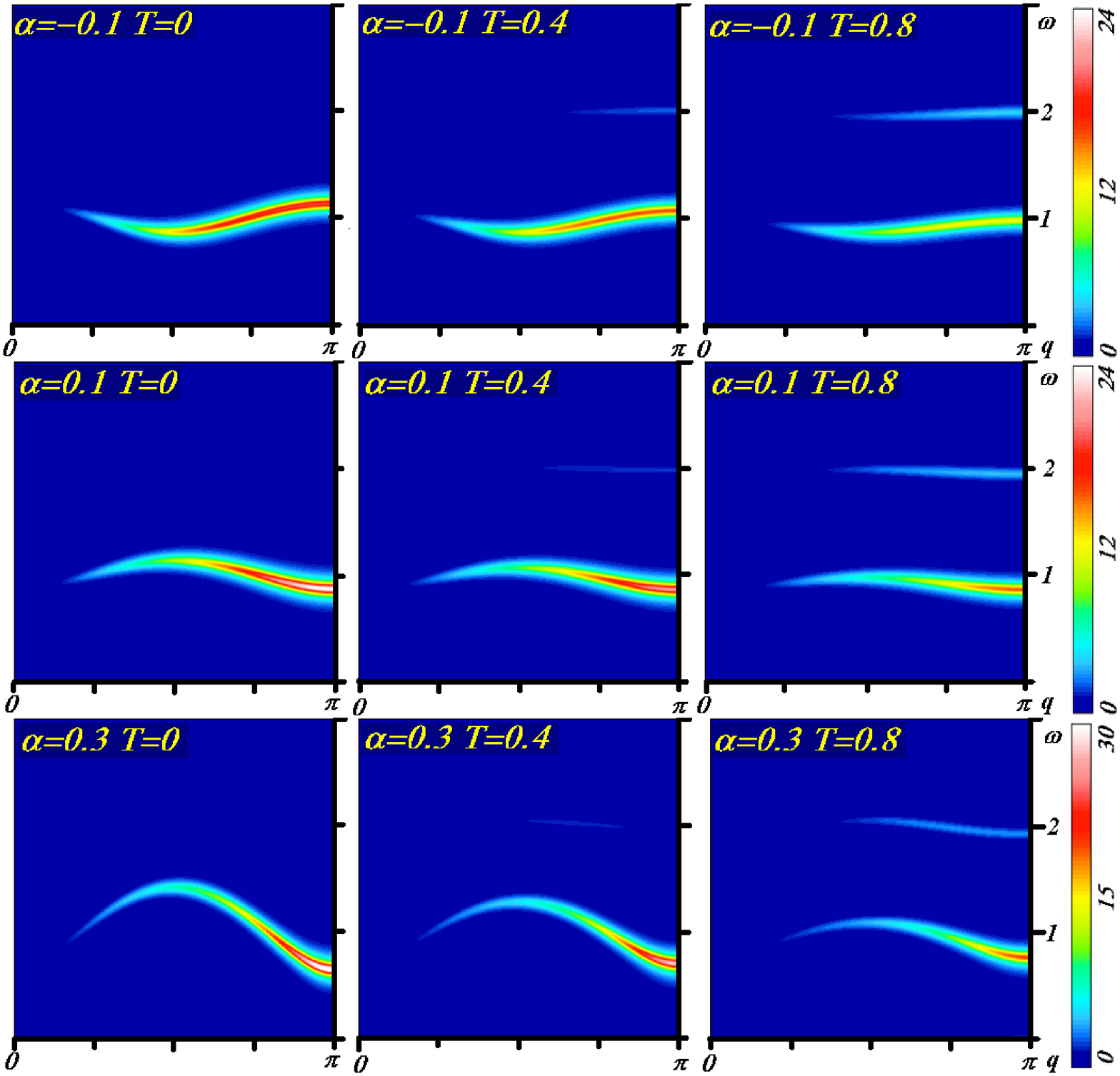}
\vspace{1mm}
\caption{$S(q,\omega)$ at $T=0, 0.4$, and $0.8$ for $\alpha=-0.1, 0.1$, and $0.3$. }
\label{fig:3-DSF}
\end{center}
\end{figure}
Using numerical results for $\lambda$ and $\omega_{\pm}(q)$, we calculate dynamical structure factors at finite temperatures. The results are shown in Fig. \ref{fig:3-DSF}. According to the finite-size analysis for the $S(q,\omega)$ obtained by the exact diagonalization method, the lowest excited states in $\alpha=-0.1$, $0.1$, and $0.3$ form the one-magnon mode in $0 \leq q \leq \pi$ \cite{me1}. 

Since we turn our attention to the behavior of the isolated mode located around $\omega \sim 1.0$ and $2.0$, we set $T<1.0$. 
At $T=0$ the intensity in $\alpha=-0.1$ takes the maximum at $q=0.95\pi$, while in $\alpha=0.1$ and $0.3$ the intensity at $T=0$ takes the maximum at $q=\pi$. We have confirmed that the wave number of the maximum intensity at $T=0$ shifts towards $\pi/2$, as $\alpha$ decreases in $\alpha<0$. The results are consistent with those obtained by the exact diagonalization method \cite{me1}.

When temperature increases, the intensity of the one-magnon mode around $\omega \sim 1.0$, which corresponds to the singlet-triplet excitation, is reduced. 
At finite temperatures, the additional mode with weak intensity emerges around $\omega \sim 2.0$. At $T=0.8$, its intensity takes the maximum at $q \sim \pi$ in $\alpha=0.1$ and at $q=0.78\pi$ in $\alpha=0.3$, respectively, whereas the intensity of the singlet-triplet mode in $\alpha>0$ takes the maximum at $q=\pi$ even at finite temperatures.
The excitation mode around $\omega \sim 2.0$ is caused by the triplet-quintet excitation in thermal equilibrium, which vanishes at $T=0$, because $\rho_{\rm t} \rightarrow 0$ and $\rho_{\rm q} \rightarrow 0$ as $T \rightarrow 0$.

As shown in Fig. \ref{fig:1-disp}, the excitation continuum is located around $\omega \sim 2.0$ in $\alpha =0.3$ \cite{me1}. Accordingly, the triplet-quintet mode becomes unstable in the continuum and disappears. In $\alpha =0.1$, by contrast, the excitation continuum is located in the higher energy region \cite{me1}. Therefore, the triplet-quintet mode may be stable. 
Recently, it was determined experimentally that NDOAP is well described by the isotropic $S=1$ bond-alternating Heisenberg chain with $\alpha \sim 0.1$ \cite{naru4}. 
When inelastic neutron-scattering experiments are performed on NDOAP, such a triplet-quintet mode with weak intensity may be observed around $\omega \sim 2.0$ in addition to the noticeable singlet-triplet mode around $\omega \sim 1.0$. According to our results, the ratio of the intensity of the higher-energy mode to that of the lower-energy mode at $q=\pi$ is evaluated to be $22 \%$ in $T=0.8$.

\section{Summary}
We have investigated dynamical properties of $S=1$ bond-alternating spin chains in the dimer phase at finite temperatures, using Pair-DCEFA. At low temperatures, the delta-function-type peak of the excitation from the singlet to triplet states appears in $\omega \sim 1$. When temperature increases, the additional weak peak caused by the excitation from the triplet to quintet states emerges in $\omega \sim 2$. This additional mode may be observed in the system with small $\alpha$ such as NDOAP.

\section*{Acknowledgments}
Numerical computations were partly carried out at the Supercomputer Center, the Institute for Solid State Physics, University of Tokyo. 
This work was supported by a Grant-in-Aid for Scientific Research from the Ministry of Education, Culture, Sports, Science, and Technology, Japan.


\end{document}